\pdfoutput=1

\documentclass[11pt]{article}

\usepackage[preprint]{acl}
\usepackage{dirtytalk}
\usepackage{comment}

\usepackage{times}
\usepackage{latexsym}

\usepackage{enumitem}

\setlist[itemize]{label=--}

\usepackage[T1]{fontenc}

\usepackage{xcolor}
\usepackage{soul}

\usepackage[utf8]{inputenc}

\usepackage{microtype}

\usepackage{inconsolata}

\usepackage{graphicx}
\usepackage{subcaption}
\usepackage{tikz}
\usetikzlibrary{arrows, positioning}

\tikzstyle{blockHR}  = [rectangle, draw=red!50,   fill=red!20,   thick, text centered,
                        rounded corners, minimum height=2em, minimum width=2em]
\tikzstyle{blockCLM} = [rectangle, draw=blue!50,  fill=blue!20,  thick, text centered,
                        rounded corners, minimum height=2em, minimum width=2em]
\tikzstyle{blockCOV} = [rectangle, draw=orange!50,fill=orange!20,thick, text centered,
                        rounded corners, minimum height=2em, minimum width=2em]
\tikzstyle{blockD}   = [rectangle, draw=green!50, fill=green!20, thick, text centered,
                        rounded corners, minimum height=2em, minimum width=2em]
\tikzstyle{blockLLM} = [rectangle, draw=yellow!50,fill=yellow!20,thick, text centered,
                        rounded corners, minimum height=2em, minimum width=2em]
\tikzstyle{blockGRY} = [rectangle, draw=gray!50,  fill=gray!20,  thick, text centered,
                        rounded corners, minimum height=2em, minimum width=2em]

\title{Towards Robust Legal Reasoning: Harnessing Logical LLMs in Law
}

\author{
  \textbf{Manuj Kant\textsuperscript{1}}\thanks{Equal Contribution},
  \textbf{Sareh Nabi\textsuperscript{2}}\footnotemark[1]\thanks{Work done outside Amazon},
  \textbf{Manav Kant\textsuperscript{3}}\footnotemark[1],\\ \\
  \textbf{Roland Scharrer\textsuperscript{4,5}}\footnotemark[1],
  \textbf{Megan Ma\textsuperscript{5}},
  \textbf{Marzieh Nabi\textsuperscript{5, 6}}\footnotemark[1]\thanks{Corresponding Author: \href{mailto:mnabi@paxai.co}{mnabi@paxai.co}}
\\  
\\
\small{
  \textsuperscript{1}Chabot-Las Positas, 
  \textsuperscript{2}Amazon,
  \textsuperscript{3}Caltech,
  \textsuperscript{4}Stanford Computer Science,
  \textsuperscript{5}Stanford CodeX,
  \textsuperscript{6}PaxAI
  }
}

\begin{document}
\maketitle

\raggedbottom
\begin{abstract}

Legal services rely heavily on text processing. While large language models (LLMs) show promise, their application in legal contexts demands higher accuracy, repeatability, and transparency. Logic programs, by encoding legal concepts as structured rules and facts, offer reliable automation, but require sophisticated text extraction. We propose a neuro-symbolic approach that integrates LLMs’ natural language understanding with logic-based reasoning to address these limitations.

As a legal document case study, we applied neuro-symbolic AI to coverage-related queries in insurance contracts using both closed and open-source LLMs. While LLMs have improved in legal reasoning, they still lack the accuracy and consistency required for complex contract analysis. In our analysis, we tested three methodologies to evaluate whether a specific claim is covered under a contract: a vanilla LLM, an unguided approach that leverages LLMs to encode both the contract and the claim, and a guided approach that uses a framework for the LLM to encode the contract. We demonstrated the promising capabilities of LLM + Logic in the guided approach.

\end{abstract}
\section{Introduction}
\subsection{Importance of Trustworthy Legal AI}
Legal systems rely on rigorous reasoning, explainability, and transparency to ensure fairness and accountability. Unlike many other AI applications, legal decision-making directly affects individuals' rights, obligations, and access to justice. Consequently, AI-driven legal solutions must go beyond surface-level predictions and provide structured, interpretable reasoning.

Expert attorneys engage in complex reasoning beyond pattern recognition. Legal analysis requires System 2 thinking — deliberate and logical reasoning that evaluates statutes, case law, and contracts. Attorneys dissect legal texts, identify principles, and construct arguments based on precedent. Their decisions involve weighing interpretations, assessing nuances, and considering broader implications. Additionally, legal professionals must articulate their reasoning clearly, ensuring their conclusions are defendable against scrutiny from courts, clients, and the opposition. 

The sensitive nature of legal queries requires a system that is both correct and interpretable. In the U.S., oversight of AI systems is intensifying, with the \citet{BipartisanAI2024} highlighting the need for transparency to prevent deceptive practices and ensure consumer protection. Every legal argument must reference laws, precedents, or contractual clauses, ensuring accountability. Unlike black-box AI, legal reasoning must be auditable, allowing stakeholders to trace conclusions. Without this level of explainability, AI legal tools risk undermining trust and reliability in decision-making.

In parallel, sector-specific supervision in the insurance domain, as in our case study, is evolving; for example, the \citet{IAIS2024} recently published its Draft Application Paper on the Supervision of AI, which calls for rigorous auditability and interpretability standards for AI-driven contract analytics. Under Europe's General Data Protection Regulation, data subjects must be provided “meaningful information about the logic” underlying automated decision-making processes \cite{GDPR2016}. This requirement ensures that individuals can understand, challenge, or seek human intervention regarding algorithmic decisions. Similarly, the proposed EU AI Act mandates that high-risk AI systems be designed with explainability and traceability, ensuring stakeholders can reasonably comprehend the system’s functioning and outputs \cite{EUAIAct2023}. 

As AI increasingly integrates into legal workflows, the need for trustworthy solutions that embody human-like reasoning, transparency, and explainability becomes more critical. AI must assist in analyzing legal texts and provide justifications that align with established legal reasoning practices. The challenge lies in designing AI systems that generate plausible answers and engage in structured, interpretable decision-making, ensuring they can be trusted in high-stakes legal contexts.

\subsection{Challenges in Legal Text Processing}
Legal services rely mainly on text-processing capabilities, which can enormously benefit from new advancements in large language models (LLM). Several scientific studies and business initiatives have highlighted the potential and limitations of LLMs in the legal domain. Nevertheless, LLM hallucinations have manifested in critical errors, such as generating nonexistent case law citations and misinterpreting contractual provisions. 

A prominent example is \textit{Mata v. Avianca}, where an attorney unknowingly submitted a brief containing fictitious judicial opinions produced by ChatGPT \citep{Aidid2024}. This event underscores the risks of using LLMs without robust verification mechanisms. 

Applying LLMs in the legal domain demands higher accuracy, repeatability, and transparency to achieve a transformative impact. The LLM reasoning abilities have traditionally been too weak to understand the complex logic associated with legal contracts. Considerable progress is still required before these technologies deliver consistent and transparent solutions. 

While human lawyers can articulate the reasoning behind their decisions and strategies, LLMs lack this capability to a sufficient degree. Despite progress in methodologies such as retrieval augmented generation - which guides LLMs to retrieve information from credible sources - hallucinations can and do occur, including for citations in the legal domain \cite{magesh-2024}. The auto-regressive nature of these models, which pushes them into greedily generating responses word-by-word rather than upfront planning, may contribute to this limitation \cite{boraz-2024}. 

The recent release of OpenAI o1, which achieved substantially better results on reasoning-based tasks than its predecessors, can change this situation. The subsequent releases of DeepSeek R1 and OpenAI o3-mini, which achieved similar results to OpenAI o1 at substantially lower costs, have demonstrated the potential for \say{reasoning} LLMs to revolutionize task automation. 

Despite these advancements, LLMs (including reasoning models such as OpenAI o1) still have a penchant for hallucinating on tasks that involve applying and interpreting complex rules. OpenAI o1 achieved a score of $77.6\%$ on LegalBench \cite{legalbench-2023, vals-2024}, a benchmark comprising a diverse set of tasks on various legal domains, leaving much room for improvement. 

\subsection{Proposed Neuro-Symbolic Approach}
Unlike LLMs, logic programs, which have proven helpful for formally representing legal concepts as structured code, offer a solution to this ambiguity by reliably automating legal reasoning. Since logic programming fundamentally relies on the interplay of rules and facts, developing computable legal reasoning may depend on a complex information extraction process from written documents \cite{wang-pan-2020, aitken-2002}. 

A neuro-symbolic AI approach of combining LLMs' natural language capabilities with a logic-based reasoning system could eventually offset LLMs' limiting drawbacks to achieve correct, consistent, and explainable text analysis, generation, and manipulation of legalese. Applying this approach raises new questions about a) architecture - how to combine LLMs with logic programs, b) performance - what is the improvement in accuracy and consistency, and c) explainability - is the reasoning more understandable for humans, compared to plain vanilla LLMs. 

This paper demonstrates how integrating LLMs with logic programming, particularly by prompting LLMs on legal terms transformed into logic programs, could outperform vanilla LLMs on targeted legal queries. Furthermore, we evaluate the performance gain by measuring the effect of prompting LLMs on legal terms transformed into logic programs compared to applying solely LLMs to query specific legal cases.

The described experiments are based on a predefined and validated set of insurance claim coverage questions and answers from two US health insurance policies: $1)$ a simplified Chubb Hospital Cash Benefit Policy (see Appendix \ref{app:simplified_chubb}) and $2)$ more complex, a Stanford Cardinal Care Aetna Student Health Insurance Plan \cite{aetna2024}.

We tested three approaches. In the \textit{vanilla LLM} approach, LLMs answered coverage questions without any guidance on how to derive the answers. In the \textit{unguided} approach, different LLMs were tasked with converting the insurance contract and claims into logic encoding (Prolog). We then determined claim coverage by conducting manual evaluations and utilizing a Prolog interpreter (SWISH). Finally, in the \textit{guided} approach, we provided the LLM with a structured framework containing basic facts and information necessary for logic encoding.

\section{Related Work}
\subsection{Evaluation of LLM in the Legal Domain}
Recent evaluations of LLMs in the legal domain have revealed promising advances and critical limitations. \citet{blairstanek2025llms} show that state-of-the-art LLMs exhibit considerable output instability when answering legal questions, with models yielding divergent decisions even under controlled settings. In parallel, \citet{hu2025finetuning} address the prevalent issue of hallucinations in legal question answering by proposing a fine-tuning framework that integrates behavior cloning with a sample-aware iterative direct preference optimization strategy, thereby enhancing factual consistency. \citet{peoples2025ai} further underscores that, although LLMs are capable of performing basic legal analysis through a typical chain of thoughts approach such as Issue, Rule, Analysis, and Conclusion (IRAC), their brief and sometimes unreliable outputs raise concerns regarding their adequacy for high-stakes legal reasoning and education. 

Complementing these findings, an evaluation reported in the \textit{Journal of Legal Analysis} \cite{oup2025} highlights persistent transparency, ethical compliance, and reliability challenges when deploying LLMs for legal research in practice. The heterogeneous nature of legal language across different jurisdictions often leads to inconsistencies in model outputs, thereby questioning the ability to generalize with the necessary level of accuracy. The study demonstrated that legal hallucinations are pervasive and disturbing: hallucination rates range from 59\% to 88\% in response to specific legal queries. 

Comprehensively the \textit{LegalBench} benchmark introduced by Guha et al. \cite{legalbench-2023} provides a collaboratively built suite of tasks that systematically measures various facets of legal reasoning, emphasizing the necessity for domain-specific evaluation metrics. These studies illustrate that while LLMs hold potential for legal applications, careful and targeted methodological improvements are essential to ensure their dependable integration into legal practice. Thus, while reported accuracy metrics are encouraging, they must be evaluated alongside limitations in consistency and transparency to assess the actual applicability of LLMs in the legal domain.




\subsection{Advances in Neuro-Symbolic AI}
Recent advances in legal language processing have increasingly focused on integrating LLMs with symbolic reasoning to balance the flexibility of neural architectures with the rigor of formal logic. \citet{ssrn5023212} demonstrated that embedding logical rules into neural frameworks can enhance the interpretability and robustness of legal text analysis. \citet{servantez2024} introduced the \textit{Chain of Logic} prompting method, which decomposes legal reasoning into independent logical steps and recomposes them to form coherent conclusions for rule‐based legal evaluation. Similarly,  \citet{InsurLE} presented \textit{InsurLE}. This domain-specific controlled natural language codifies insurance contracts by preserving key syntactic nuances while exposing the underlying formal logic for a computable representation. 

\citet{wei2005} proposed a hybrid neural-symbolic framework that synergizes neural representations with explicit logical rules, thereby improving the rigor of legal reasoning in automated systems. \citet{patil2025} systematically surveyed methods to enhance reasoning in LLMs and highlighted modular reasoning and retrieval-augmented techniques as promising approaches for bolstering logical consistency in legal applications. \citet{colelough2025} provided a comprehensive review of neuro-symbolic AI in the legal domain, identifying substantial progress in learning and inference while noting significant gaps in explainability and understanding derived logic programs. 

\citet{calanzone2024} developed a neuro-symbolic integration approach that enforces logical consistency by incorporating external constraint sets into LLM outputs. \citet{sun2024} introduced a framework that explicitly learns case-level and law-level logic rules to generate faithful and interpretable explanations for legal case retrieval. \citet{tan2024} enhanced LLM reasoning through a self-driven Prolog-based chain-of-thought mechanism that iteratively refines logical inferences in legal tasks. Lastly, \citet{Vakharia_2024} proposed \textit{ProSLM}, a Prolog-based language model that validates LLM outputs against a domain-specific legal knowledge base, ensuring higher factual accuracy and interpretability in legal question answering.

Collectively, these studies chart a clear trajectory toward AI systems that harness the complementary strengths of deep learning and logical inference to address the nuanced challenges inherent in legal reasoning.

\section{Preliminary Experiments}
\label{sec:exp_vanilla_unguided}
In this section, we evaluate a range of state-of-the-art reasoning models to benchmark their capabilities in answering coverage-related `yes/no' claim questions about an insurance policy.

We selected seven LLMs, including O1-preview, DeepSeek-R1, Llama-3.1-405B-Instruct, Claude-3.5-Sonnet, Mistral-Large-Latest, Gemini-1.5-Pro, and GPT-4o-2024-08-06\footnote{We included GPT-4o-2024-08-06 along with all aforementioned state-of-the-art reasoning models to assess its performance in legal and contract analysis tasks, given its strong contextual comprehension and broad reasoning ability.}. These models excel in long-context reasoning, mathematical problem-solving, multi-step reasoning, logical consistency, and following policy rules, making them well-suited for analyzing insurance contracts and legal texts. For all models, we set both the temperature and top-p parameters to $1$.

The insurance contract used in the experiments in this section is the Simplified Chubb Hospital Cash Benefit policy, referred to as Chubb hereafter. The task is to determine whether nine given claims are covered under this insurance policy. The Chubb contract and nine claim queries are provided in Appendix \ref{app:simplified_chubb} and  \ref{app:queries_and_answers}, respectively.

We describe the vanilla LLM approach in \S\ref{sec:vanilla_llm} where we directly ask the LLM to answer `yes/no' claim questions. In Section \ref{sec:exp_unguided}, we task the LLM with generating Prolog encodings of the insurance contract and claim queries, which we then manually evaluate using the help of SWISH Prolog interpreter \cite{swish}.

\subsection{Vanilla LLM Approach} 
\label{sec:vanilla_llm}
We prompt the selected LLMs to answer nine claim questions about the Chubb insurance policy. We then evaluate the performance of these models across 10 trials and report their average accuracies and standard errors in Figure~\ref{fig:vanilla_unguided_chubb}. The prompt used for the LLMs is provided in Appendix \ref{app:prompt_vanilla}.

The results show that models such as Mistral-large-latest, Gemini-1.5-pro, Claude-3.5-sonnet, Llama-3.1-405B-instruct, and GPT-4o-2024-08-06 achieved a consistent accuracy of $0.78$ across all 10 trials, with no variance, even at a temperature setting of $1.0$. The error bars in Figure~\ref{fig:vanilla_unguided_chubb} represent the Standard Error of the Mean (SEM)\footnote{The SEM is calculated by dividing the standard deviation of accuracy scores from 10 trials by the square root of the number of trials.}, indicating the variability of the model across trials.

All models consistently failed to correctly answer questions 5 and 9 (see Appendix \ref{app:queries_and_answers}). Question 5 asks whether a self-harm injury is covered if all other conditions are met, while Question 9 concerns coverage for a police officer injured outside of duty (see Appendix \ref{app:queries_and_answers} for the exact wording). Clause 1.1 of the policy specifies that hospitalization must result from sickness or accidental injury, meaning the claim in Question 5 is not covered. In Question 9, although "Service in the police" is excluded if the injury arises from it, the injury in this case occurred when the officer’s son bit him in the ankle outside of duty. In both cases, the vanilla LLM models struggled to distinguish between being a police officer and being injured outside of service, as well as failing to recognize that punching someone in the face is not classified as sickness or accidental injury.

In contrast, DeepSeek-R1 achieved an average accuracy of $0.81$ with an SEM of $0.02$, correctly answering Question 5 in 3 out of 10 trials, though it still missed Question 9 in all trials. The O1-preview model performed better, achieving an average accuracy of $0.88$ with an SEM of $\pm0.02$. It correctly answered Question 9 in 9 out of 10 trials but failed to answer Question 5 correctly in 9 out of 10 trials.

As observed, the vanilla LLM approach alone cannot provide answers to claim questions with 100\% accuracy and consistency across trials, even when using state-of-the-art reasoning models. Next, we aim to enhance LLMs by leveraging the benefits of logic programming. We will ask them to generate Prolog encodings of the policy and claims and then answer the claim questions by evaluating the generated Prolog encodings.
\begin{figure*}[ht]
    \centering
    \begin{minipage}{1\textwidth}
    \centering                  \includegraphics[width=0.77\textwidth]{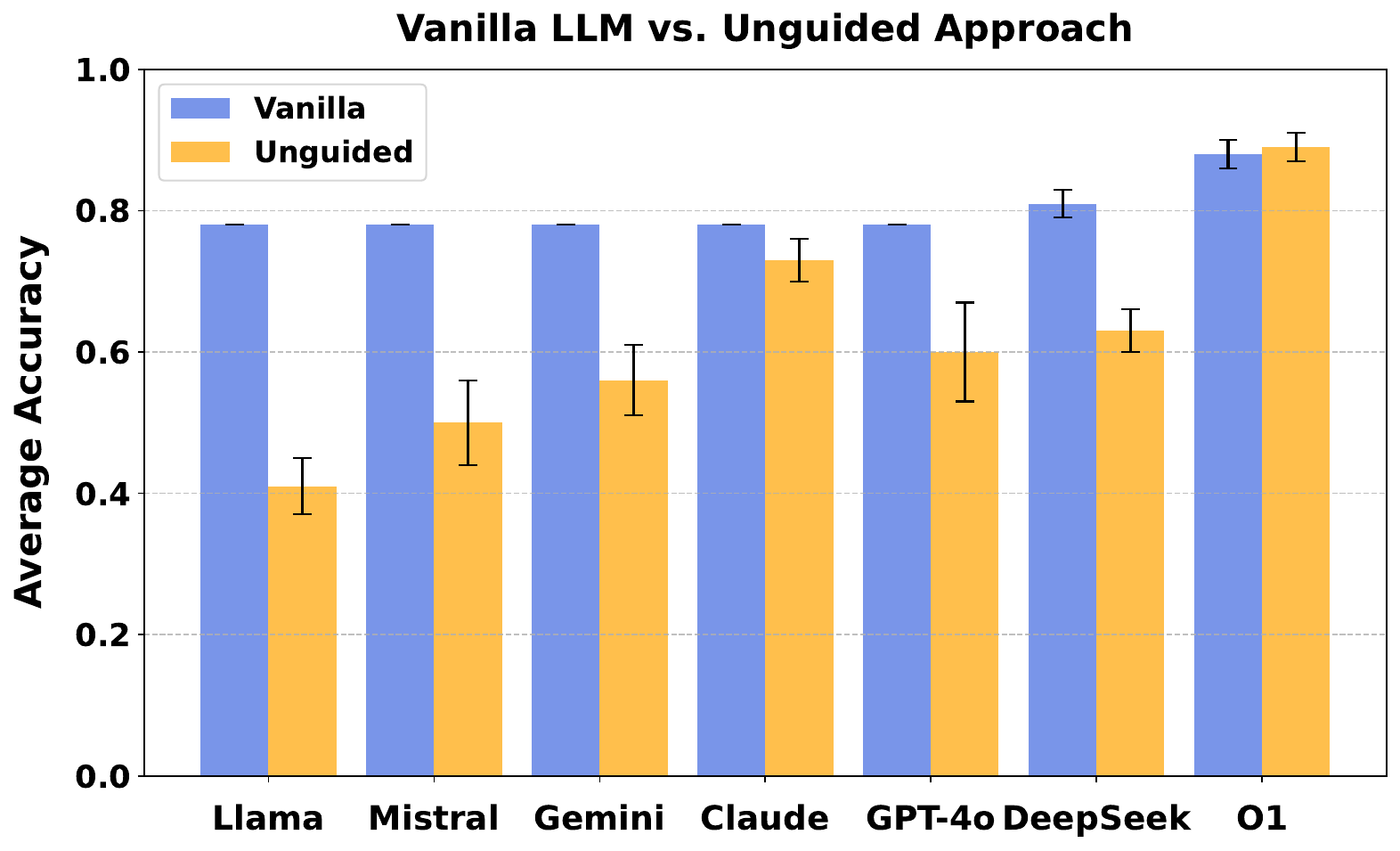}
    \end{minipage}
    
    \begin{minipage}{0.48\textwidth}
    \centering
    \begin{tabular}{|c|c|}
    \hline
    \multicolumn{2}{|c|}{\textbf{Vanilla LLM Approach}} \\ \hline
    Model & Accuracy $\pm$ SEM \\ \hline
    Mistral-large-latest & 0.78 $\pm$ 0.00 \\ \hline
    Gemini-1.5-pro & 0.78 $\pm$ 0.00 \\ \hline
    Claude-3.5-sonnet & 0.78 $\pm$ 0.00 \\ \hline
    Llama-3.1-405B-instruct & 0.78 $\pm$ 0.00 \\ \hline
    GPT-4o-2024-08-06 & 0.78 $\pm$ 0.00 \\ \hline
    DeepSeek-R1 & 0.81 $\pm$ 0.02 \\ \hline
    O1-preview & 0.88 $\pm$ 0.02 \\ \hline
    \end{tabular}
    \end{minipage}
    \hfill
    \begin{minipage}{0.48\textwidth}
    \centering
    \begin{tabular}{|c|c|}
    \hline
    \multicolumn{2}{|c|}{\textbf{Unguided Approach}} \\ \hline
    Model & Accuracy $\pm$ SEM \\ \hline
    Mistral-large-latest & 0.50 $\pm$ 0.06 \\ \hline
    Gemini-1.5-pro & 0.56 $\pm$ 0.05 \\ \hline
    Claude-3.5-sonnet & 0.73 $\pm$ 0.03 \\ \hline
    Llama-3.1-405B-instruct & 0.41 $\pm$ 0.04 \\ \hline
    GPT-4o-2024-08-06 & 0.60 $\pm$ 0.07 \\ \hline
    DeepSeek-R1 & 0.63 $\pm$ 0.03 \\ \hline
    O1-preview & 0.89 $\pm$ 0.02 \\ \hline
    \end{tabular}
    \end{minipage}
    
    \caption{LLM models' average accuracy on the Chubb insurance claim coverage dataset. 
    The plot (top) visualizes the models' average accuracy with error bars representing the Standard Error of the Mean (SEM) across 10 trials. 
    The tables (bottom) provide the corresponding raw numerical accuracy values: the left table represents the Vanilla LLM approach, while the right table corresponds to the Unguided Prolog Generation approach.}    
    \label{fig:vanilla_unguided_chubb}
\end{figure*}

\subsection{Unguided LLM-generated Prolog}
\label{sec:exp_unguided}
We prompted the selected LLMs (from \S\ref{sec:exp_vanilla_unguided}) to generate Prolog encodings of the Chubb policy contract and nine claims. We then manually evaluated whether the insurance covered the claims by evaluating the Prolog encodings and using the help of SWISH Prolog interpreter\footnote{The generated encodings often failed to run on SWISH due to its ambiguity. In such cases, we manually reasoned through the policy encodings and evaluated the claims.}. 

This process was repeated for 10 trials. In every trial, each LLM generated a policy encoding from the prompt in Appendix~\ref{app:prompt_unguided}, and then translated nine claim questions (given in \ref{app:queries_and_answers}) into Prolog queries based on the policy encoding. We manually evaluated the policy and claim encodings and recorded the number of correct responses. When the encodings were unambiguous, we confirmed our evaluation using SWISH.
We report average accuracies and the standard error of the mean over these 10 trials in Figure~\ref{fig:vanilla_unguided_chubb} and provide a qualitative analysis of each LLM for this task below.

The O1-preview model achieved an average accuracy of $0.89 \pm 0.02$, slightly improving on its vanilla approach in \S\ref{sec:vanilla_llm}. O1-preview answered Question 9 correctly in all trials, showing improved reliability over the vanilla approach in distinguishing between an injury caused by a son biting the claimant while the claimant was a police officer—an explicitly covered scenario. 

However, similar to its vanilla approach, the O1-preview struggled with Question 5, missing it in 9 out of 10 trials. This question involved self-harm, where the claimant was injured due to a face punch. O1-preview failed to determine whether the scenario fell under an exclusion correctly. While it correctly excluded activities like skydiving, firefighting, and police service (where injuries during these activities are not covered), it failed to identify the primary cause of the claim—whether it was sickness or accidental injury, which are addressed indirectly in the contract. Additionally, O1-preview missed Question 4 in only one trial due to ambiguity in encoding time-based conditions. The vanilla LLM, in comparison, had a slightly lower average accuracy of $0.88 \pm 0.02$, with errors mostly in Question 5, and it consistently failed to answer Questions 9 and 4. 

The DeepSeek-R1 model achieved an average accuracy of $0.63 \pm 0.03$. In several trials, it generated incorrect logic encodings, particularly in how it handled exclusions. For instance, it sometimes treated exclusions as conjunctive conditions (e.g., both general activity exclusions, such as serving as a firefighter, and the age > 80 condition had to hold simultaneously). However, the policy contract (see Appendix \ref{app:simplified_chubb}) specifies that exclusions should be interpreted with an OR operator: coverage is denied if sickness or accidental injury results from a listed activity (e.g., skydiving, military service) or if the claimant is 80 years or older at the time of hospitalization. This misinterpretation led to inaccuracies in claim assessments. Some queries were ambiguous, preventing DeepSeek-R1 from determining a final coverage decision. Compared to O1-preview, which achieved $0.89 \pm 0.02$, DeepSeek-R1 not only had a lower average accuracy ($0.63 \pm 0.03$) but also exhibited more significant variability across trials.


GPT-4o-2024-08-06 achieved an average accuracy of $0.60 \pm 0.07$. GPT-4o demonstrated errors in encoding policy rules in some trials, and its claim encodings often lacked sufficient information. This led to ambiguity, preventing the determination of essential predicate values required for accurate evaluation. Frequently, a final answer could not be determined due to this ambiguity. Additionally, the model exhibited significant inconsistencies across trials, producing logic encodings of varying quality. In some cases, inaccurate encodings—such as confusion between the claim date and the wellness visit time limit or inconsistent predicate parameters (e.g., passing the claim date instead of the claimants' age to an age exclusion predicate; see Appendix \ref{app:simplified_chubb})—led to a low accuracy of 1 out of 9 correct answers in one trial, while in another, it correctly answered 8 out of 9 questions.

The encodings of the remaining models were often ambiguous, resulting in inaccurate responses to several questions. The Mistral-large-latest model had an average accuracy of $0.5 \pm 0.06$, with variability and significant struggles in determining coverage due to ambiguous logic encodings. The Gemini-1.5-Pro model also had an average accuracy of $0.56 \pm 0.05$ and faced logic ambiguity issues. 
The Claude-3.5-Sonnet model achieved an average accuracy of $0.73 \pm 0.03$, with its main challenges primarily in Questions 5 and 9. Finally, the Llama-3.1-405B-instruct model, with an average accuracy of $0.41 \pm 0.04$, faced frequent ambiguities. All these models performed worse than their vanilla versions, which had a consistent accuracy of $0.78$. Overall, the ambiguity and inaccuracy in their encodings led to challenges in accurately responding to claim queries. As the next step, we propose expert-guided Prolog encoding generation in \S\ref{sec:expert-informed} to improve accuracy and consistency in LLMs when answering such claims.

\raggedbottom
\section{Expert-Guided Experiments}

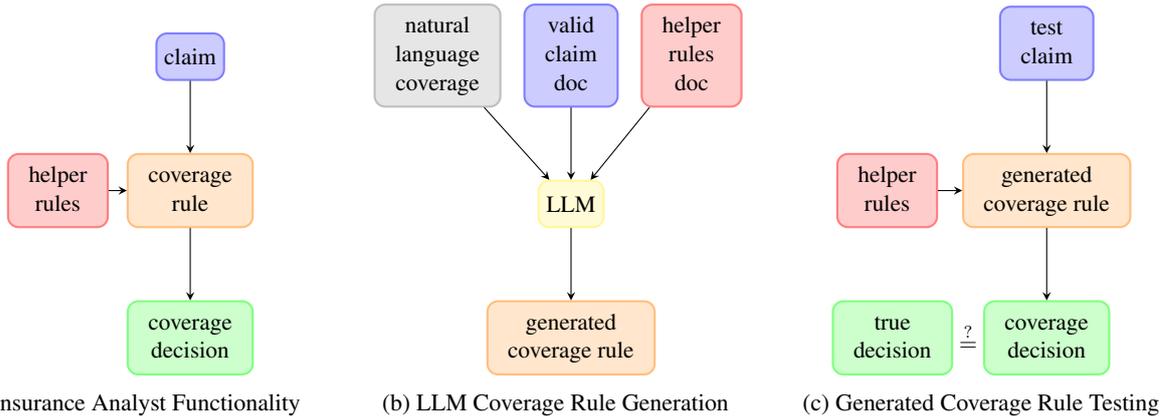
\begin{figure*}[t]
\centering
%
%
\begin{subfigure}{0.3\textwidth}
\centering
\begin{tikzpicture}[node distance=1.2cm,>=stealth,auto,
                    scale=0.8,transform shape]

\node[blockHR] (helper)
  {\begin{tabular}{c}
   helper\\rules
   \end{tabular}};
\node[blockCOV, right=.3cm of helper] (coverage)
  {\begin{tabular}{c}
   coverage\\rule
   \end{tabular}};
\node[blockD,   below=of coverage] (decision) {\begin{tabular}{c}
   coverage\\decision
   \end{tabular}};
\node[blockCLM, above=of coverage] (claim)   {claim};

\draw[->] (helper) -- (coverage);
\draw[->] (claim)  -- (coverage);
\draw[->] (coverage) -- (decision);

\end{tikzpicture}
\caption{Insurance Analyst Functionality}
\label{fig:insurance-analyst-functionality}
\end{subfigure}
\hfill
%
%
\begin{subfigure}{0.3\textwidth}
\centering
\begin{tikzpicture}[node distance=1.2cm,>=stealth,auto,
                    scale=0.8,transform shape]

\node[blockLLM] (llm) {LLM};

\node[blockGRY, above left=1.2cm and 0.6cm of llm] (nlc)
  {\begin{tabular}{c}
   natural\\language\\
   coverage
   \end{tabular}};
\node[blockCLM, above=1.2cm of llm] (valid)
  {\begin{tabular}{c}
   valid\\claim\\
   doc
   \end{tabular}};
\node[blockHR, above right=1.2cm and 0.6cm of llm] (hrdoc)
  {\begin{tabular}{c}
   helper\\rules\\
   doc
   \end{tabular}};

\node[blockCOV, below=of llm] (genCov)
  {\begin{tabular}{c}
   generated\\
   coverage rule
   \end{tabular}};

\draw[->] (nlc)   -- (llm);
\draw[->] (valid) -- (llm);
\draw[->] (hrdoc) -- (llm);
\draw[->] (llm)   -- (genCov);

\end{tikzpicture}
\caption{LLM Coverage Rule Generation}
\label{fig:llm-coverage-rule-generation}
\end{subfigure}
\hfill
%
%
\begin{subfigure}{0.3\textwidth}
\centering
\begin{tikzpicture}[node distance=1.2cm,>=stealth,auto,
                    scale=0.8,transform shape]

\node[blockCLM] (testclaim)
  {\begin{tabular}{c}
   test\\
   claim
   \end{tabular}};
\node[blockCOV, below=of testclaim] (genCov)
  {\begin{tabular}{c}
   generated\\
   coverage rule
   \end{tabular}};
\node[blockHR, left=.4cm of genCov] (helper)
  {\begin{tabular}{c}
   helper\\
   rules
   \end{tabular}};
\node[blockD, below=of genCov] (decision)
{\begin{tabular}{c}
   coverage\\decision
   \end{tabular}};
\node[blockD, left=.5cm of decision]
  (correct) {\begin{tabular}{c}true\\decision\end{tabular}};

\draw[->] (testclaim) -- (genCov);
\draw[->] (helper)    -- (genCov);
\draw[->] (genCov)    -- (decision);
\draw[draw opacity=0]
  (decision.west) -- (correct.east)
  node[midway, anchor=center, yshift=.05ex] {$\stackrel{?}{=}$};

\end{tikzpicture}
\caption{Generated Coverage Rule Testing}
\label{fig:generated-coverage-rule-testing}
\end{subfigure}

\caption{Experimental overview:
  (a)~Functionality of the CodeX Insurance Analyst coverage rules.
  (b)~The LLM is prompted to generate its own version of the coverage rule given the text of the coverage and documentation of the valid claims and helper rules it can call.
  (c)~The LLM's generated coverage rule is tested by passing it test claims and determining if the correct coverage decisions were made.}
\label{fig:guided_procedure}
\end{figure*}

\label{sec:expert-informed}
In what follows, we demonstrate a workflow for leveraging LLMs \emph{through expert guidance} to automate the process of encoding health insurance policies as logic programs (also called computable contracts). 


 We prompted LLMs to encode Prolog rules representing three insurance coverages. The first coverage was the simplified Chubb policy, described in the previous experiment in \S\ref{sec:exp_vanilla_unguided} (see Appendix \ref{app:simplified_chubb}). The latter two have been derived from the Stanford CodeX \textit{Insurance Analyst} \cite{insurance-analyst-2025}, a deployed, expert-encoded computable contract representing the Stanford Cardinal Care Aetna Student Health Insurance Plan \cite{aetna2024} \cite{goodenough2023}. 
 
 Specifically, we evaluated LLMs' ability to encode the \textit{Advanced Reproductive Technology} (ART) and the \textit{Comprehensive Infertility} (CI) coverage rules from the \textit{Insurance Analyst}. Each coverage rule in the \textit{Insurance Analyst} evaluates claims to reach coverage decisions, calling helper rules from other parts of the code base in the process (see Figure \ref{fig:insurance-analyst-functionality}). The LLMs were prompted to encode their own versions of these coverage rules with $1)$ the coverage text from the Cardinal Care policy $2)$ documentation defining a valid claim to the rule, and $3)$ documentation defining the relevant helper rules which can be called from other parts of the code base (see Figure \ref{fig:llm-coverage-rule-generation}, Appendix \ref{app:art-prompt}). The documentation provided to the LLM constitutes guidance given by an expert. We had each LLM generate an encoding of each coverage $5$ times, testing each coverage encoding by querying it with claims and evaluating whether the outputted decisions were correct (see Figure \ref{fig:generated-coverage-rule-testing}).

\subsection{Guided LLM-generated Prolog for a simplified policy}
\label{sec:expert-informed-chubb}
On the Chubb policy, we prompted LLMs with the text of the policy and documentation about the facts provided in any valid claim (e.g., \verb|claim_hospitalization_reason|, \verb|claim_misrepresentation_occurred|) to be used in generating a representative computable contract. Since this policy is stand-alone, its encoding does not need to integrate into a more extensive code base. Thus, no helper rules (from other parts of the code base) needed to be included in the prompt.

Three of the four LLMs performed well on this task (see Table \ref{tab:chubb-simplified}), with each of their $5$ generated encodings perfectly answering all $9$ test queries used for evaluation. These test queries were Prolog translations of the natural language queries used to assess the previous approach (see Appendix \ref{app:queries_and_answers}). DeepSeek-R1, however, produced one encoding with a syntactic error due to an unclosed parenthesis. This, along with some failed test cases in another one of its encodings, resulted in a lower accuracy rate than the other models.

\begin{figure}[ht]
    \centering
   \includegraphics[angle=0, width=0.48\textwidth]{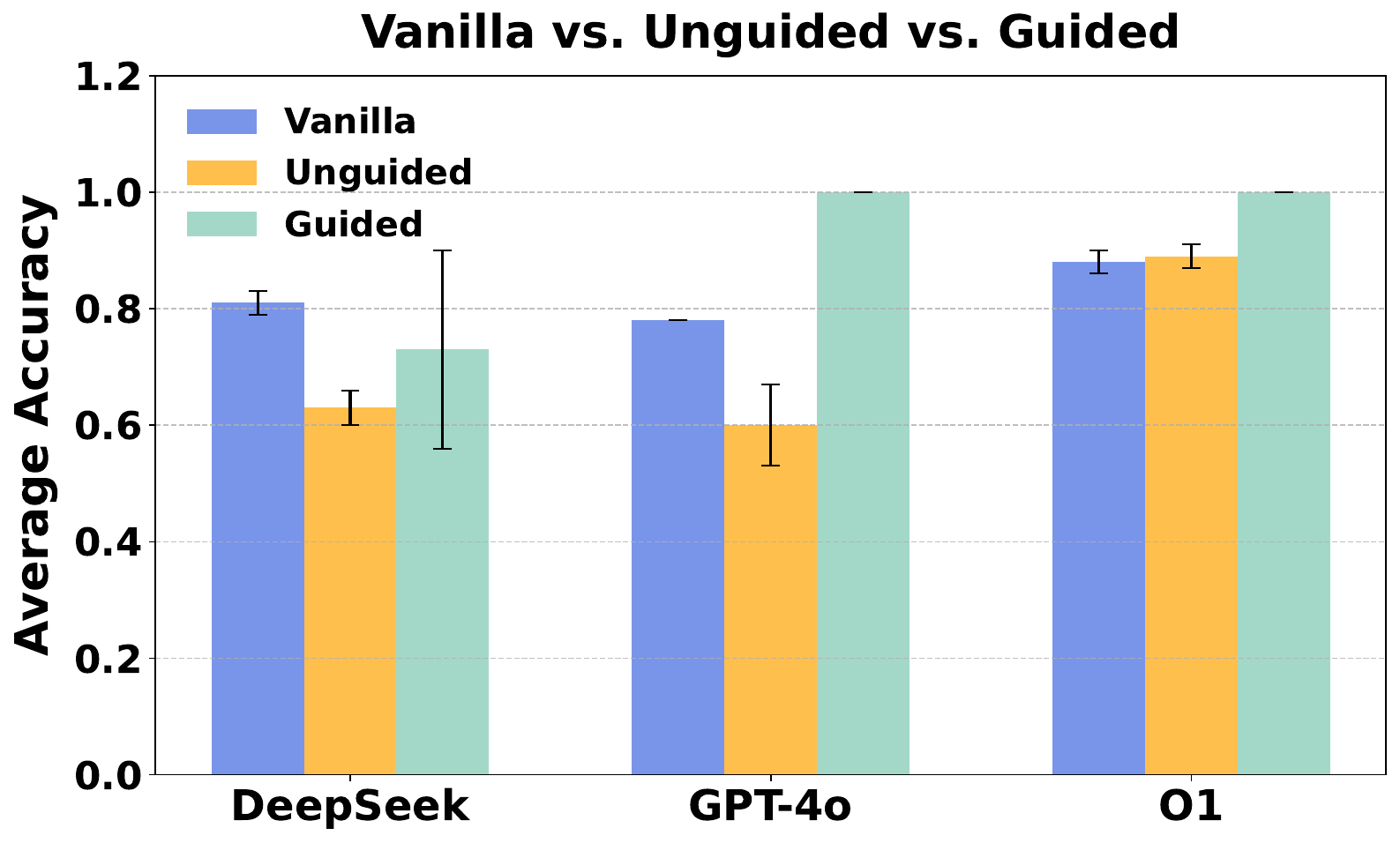}
   \caption{Average accuracy for the simplified Chubb contract across three approaches—Vanilla LLM, Unguided, and Guided—and three models, GPT-4o, o1-preview, and DeepSeek-R1, with error bars representing the standard error of the mean across 10 trials.}
\label{fig:vanilla_unguided_guided_chubb_comparisons}
\end{figure}

Figure \ref{fig:vanilla_unguided_guided_chubb_comparisons} illustrates a comparison of three approaches—Vanilla LLM, the Unguided approach described in \S\ref{sec:exp_vanilla_unguided}, and the Guided approach described in \S\ref{sec:expert-informed-chubb}—on the Chubb contract using three models: DeepSeek, GPT-4o, and OpenAI o1. As observed, both the Vanilla and Unguided approaches struggled to achieve 100\% accuracy and consistency in all models. In contrast, the Guided approach using GPT-4o and OpenAI o1 models achieved 100\% accuracy with no variations across all trials examined (zero standard error). Among the three models, OpenAI o1 performed best across all approaches.
\begin{table}
  \centering
  \begin{tabular}{|c|c|}
    \hline
    Model & Accuracy ± SEM \\
    \hline
    \verb|GPT-4o|     & $1.00 \pm  0.00$         \\ \hline
    \verb|OpenAI o1|     & $1.00 \pm  0.00$          \\ \hline
    \verb|OpenAI o3-mini|     & $1.00 \pm  0.00$          \\ \hline
    \verb|DeepSeek-R1|     & $0.73 \pm  0.17$                   \\\hline
  \end{tabular}
  \caption{Evaluation of the Guided LLM-generated logic program encodings of simplified Chubb contract.}
  \label{tab:chubb-simplified}
\end{table}
\subsection{Guided LLM-generated Prolog for coverages in a larger policy}
The Stanford Cardinal Care health insurance policy comprises many individual \say{coverages}. For a claim to be covered under the policy, it must be covered under one of these coverages. Thus, while the \textit{Insurance Analyst} has an overarching \say{covered} rule (which should be satisfied exactly when a claim is covered under the Cardinal Care policy), it also contains many rules associated with specific coverages, one of which must be satisfied for the overarching one to be satisfied. The code undergirding the \textit{Insurance Analyst}, written in Epilog (a logic programming language similar to Prolog), is available in its public code repository \cite{insurance-analyst-github-2025}. We used a Prolog-translated version of this code for consistency with our simplified Chubb experiments, asking LLMs to encode the ART and CI coverages, testing the accuracy of these encodings through $20$ test cases from the \textit{Insurance Analyst}'s publicly available code repository.

OpenAI o1 was substantially more successful at encoding these Cardinal Care coverages than GPT-4o, OpenAI o3-mini, and DeepSeek-R1. As shown in Table \ref{tab:art}, OpenAI o1's ART encodings achieved an average accuracy of $95\%$, far superseding the $50-60\%$ accuracies of the encodings generated by the other models. Similarly, as shown in Table \ref{tab:comprehensive}, OpenAI o1's encodings on CI coverages had an $87\%$ accuracy significantly outperformed that of the other models.

Since the ART and CI coverages are longer and more logically complex than the simplified Chubb policy, they serve as better differentiators of the logical capabilities of the tested LLMs. As an example of the difference in logical correctness between the logic programs written by OpenAI o1 and GPT-4o, consider the following excerpt from the ART coverage:

\begin{quote}
For women $39$ years of age and older, ovarian responsiveness is determined by measurement of day $3$ FSH obtained within the prior $6$ months. For women who are less than $40$ years of age, the day $3$ FSH must be less than $19$ mIU/mL in their most recent laboratory test to use their own eggs. For women $40$ years of age and older, their unmedicated day $3$ FSH must be less than $19$ mIU/mL in all prior tests to use their own eggs.
\end{quote}

Note that there are two age-based boundaries specified in this excerpt. Firstly, women who are at least $39$ years of age must have had an FSH test within the prior $6$ months, whereas this condition does not apply to younger women. Secondly, women who are at least $40$ will have \emph{all FSH tests} past age $40$ examined, whereas younger women will only have the most recent test looked at.

GPT-4o, in its first trial, encoded the FSH criteria in the rule \verb|validate_day_3_fsh(C)| (see Appendix \ref{app:4o-fsh}). This rule correctly checks for the strictness criterion with a boundary at age $40$, but there is no sign of the recency criterion with a boundary at age $39$. By contrast, consider the analogous encoding generated by OpenAI o1 in \emph{its} first trial of the rule \verb|meets_fsh_criteria(C)| (see Appendix \ref{app:o1-fsh}), which correctly delineates \emph{both} age-based boundaries--at age $40$ as well as $39$. Unlike the encoding generated by GPT-4o, it ensures that the most recent FSH test for women who are at least $39$ years of age was conducted no more than $6$ months ago.

This and other examples demonstrate the significant gap in logical ability between OpenAI o1 and GPT-4o, explaining the former's significantly higher accuracy in representing insurance coverages in a logical form.

OpenAI o3-mini and DeepSeek also performed worse than OpenAI o1 on ART due to logical errors and syntactical mistakes. One major issue was the misapplication of the premature ovarian failure (POF) exception. OpenAI o3-mini wrongly applied this exception to \emph{all} women aged $40+$ (not just ones with POF), allowing some to qualify when they should not have. Both models also made syntax errors that prevented their encodings from running. OpenAI o3-mini referenced the nonexistent rule \texttt{day\_3\_fsh\_tests\_since\_age\_40} where it should have been referring to \texttt{day\_3\_fsh\_tests\_since\_age\_40\_in\_claim}, while DeepSeek introduced an unclosed parenthesis, making its Prolog code invalid and thus impossible to evaluate.

Since the CI coverage is even longer and more complex than ART, three of the four models performed worse on encoding this coverage (see Table \ref{tab:comprehensive}). However, OpenAI o1 still led the pack with $87\%$ accuracy rate by producing logically and syntactically superior Prolog encodings. These results show that strong reasoning LLMs such as OpenAI o1 could play a critical role in developing computable contracts that provide reliable, interpretable, and auditable coverage decisions. 

\begin{figure*}[ht]
    \centering
   \includegraphics[angle=0, width=0.7\textwidth]{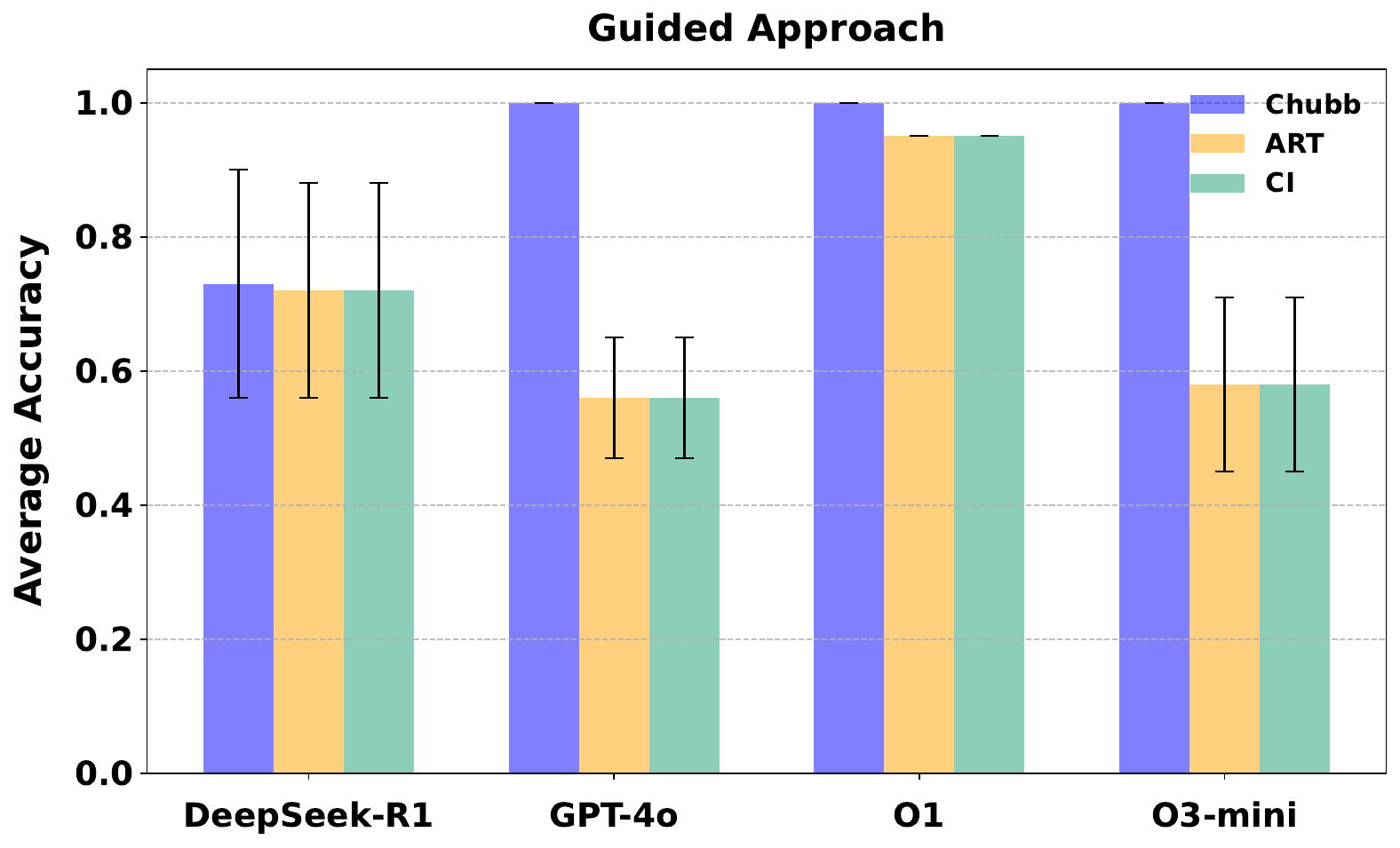}
   \caption{LLMs' average accuracy for Chubb, ART, and CI coverages using Guided approach. Error bars represent the standard error of the mean across 10 trials. Models used are Deepseek-R1, GPT-4o, OpenAI o1, and o3-mini.}
   \label{fig:guided_all_coverages_accuracy}
\end{figure*}

Finally, Figure~\ref{fig:guided_all_coverages_accuracy} consolidates the results of the Guided approach presented in Tables~\ref{tab:chubb-simplified}, \ref{tab:art}, and \ref{tab:comprehensive} for all three contracts—Chubb, ART, and CI. As shown, the OpenAI o1 model outperformed the others across all contracts, achieving 100\% accuracy on Chubb and demonstrating higher accuracy than other models on the more complex contracts, ART and CI.
\begin{table}
  \centering
  \begin{tabular}{|c|c|}
    \hline
    Model & Accuracy ± SEM \\
    \hline
    \verb|GPT-4o|     & $0.56 \pm  0.09$         \\ \hline
    \verb|OpenAI o1|     & $0.95 \pm  0.00$          \\ \hline
    \verb|OpenAI o3-mini|     & $0.58 \pm  0.13$      \\ \hline
    \verb|DeepSeek-R1|     & $0.72 \pm 0.16$
        \\\hline
  \end{tabular}
  \caption{Evaluation of the Guided LLM-generated Cardinal Care ART coverage logic program encodings.}
  \label{tab:art}
\end{table}

\begin{table}
  \centering
  \begin{tabular}{|c|c|}
    \hline
     Model & Accuracy ± SEM \\
    \hline
    \verb|GPT-4o|     & $0.37 \pm  0.10$         \\ \hline
    \verb|OpenAI o1|     & $0.87 \pm  0.04$          \\ \hline
    \verb|OpenAI o3-mini|     & $0.72 \pm  0.04$    \\ \hline
    \verb|DeepSeek-R1|     & $0.47 \pm 0.18$
        \\\hline
  \end{tabular}
  \caption{Evaluation of the Guided LLM-generated Cardinal Care CI coverage logic program encodings.}
  \label{tab:comprehensive}
\end{table}

\section{Conclusion and Future Work}
We are on the cusp of an exciting era where AI enhances legal access through human-like thinking, such as planning and reasoning. While LLMs show promise, their probabilistic nature, inconsistency, and potential for hallucination make their application in the legal domain risky. 

We propose a neuro-symbolic approach that combines LLMs with logic programming and provide a comparative analysis of a use case involving health insurance coverage questions. First, we experimented with a vanilla LLM approach, without logic programs, prompting the LLM to respond to coverage claims based on the contract. Our key observation is that advancements in foundational models allow a vanilla LLM to reasonably assess whether a claim is covered. However, it lacks full accuracy and consistency, which are essential in legal use cases.

We then prompted LLMs to convert a legal contract into a logical encoding and evaluate coverage queries with the help of a logic interpreter. In the unguided approach, where the LLM received only the contract and claim queries without additional guidance, we observed poor-quality encodings, often exhibiting ambiguity and inaccuracy, though some LLMs performed better than others. 

The results of the unguided approach were worse than those of the vanilla LLM. This poor encoding quality is expected, as these models are not specifically trained for such tasks. To address these issues, we introduced a structured framework containing essential information a human encoder would use, providing guidance to the LLMs. Our findings suggest that this guided approach significantly improves the quality, accuracy, and consistency of the generated encodings.

Beyond using LLMs for logical representations through the unguided and guided methods presented in this work, we propose exploring additional approaches in future work.

Our first proposal involves fine-tuning foundational models with high-quality, human-generated logic encodings. Generating a logic encoding for a legal segment or contract is similar to writing code. However, current foundational models have been trained on significantly more high-quality code than on logic encodings. Fine-tuning with curated logic encodings could improve LLMs’ ability to generate accurate and structured representations.

We see an opportunity to enhance the accuracy of LLM-generated Prolog encodings using agentic AI. This includes automating both LLM-generated logic encoding and evaluation within a Prolog interpreter like SWISH. Our experiments revealed frequent syntax errors in LLM-generated encodings, but this approach enables automatic error detection and correction, improving accuracy. Another approach in this direction involves using multiple LLMs: one to encode legal terms and queries, a cost-efficient model to execute them, and another to evaluate the outcomes. 
However, whether this method guarantees accurate and reliable results remains uncertain.

Our third proposal is to improve LLMs’ ability to generate accurate Prolog encodings by incorporating reinforcement learning (RL) with synthetic data and feedback from a Prolog interpreter. One potential approach is to use RL with reward modeling, where the LLM generates logical encodings, executes them in a Prolog interpreter like SWISH, and receives a reward signal based on correctness, consistency, and execution success. 

This iterative process would enable the model to refine its encoding strategy over multiple training cycles. Additionally, an RL-based framework could optimize generalization by training the model on diverse logical structures, helping it adapt to different legal contracts and reasoning tasks. Combining RL with human-in-the-loop evaluation could further enhance reliability, ensuring that generated encodings align with legal reasoning principles.

\section{Limitations}

Our current approach addresses only a limited scope and serves as an initial step in a novel direction—combining LLMs and logic programs to form a neuro-symbolic AI for legal analysis. The experiments in this work are constrained in terms of problem space, architectural design, datasets, logic interpreters used, prompt tuning, measurement metrics, and LLMs experimented with.

Our long-term ambition is to apply a neuro-symbolic approach more broadly in the legal domain. Currently, we focus only on health insurance-related coverage questions and answers. Further application areas, such as reasoning about civil and corporate legal terms, remain out of scope but present an exciting direction for future work.

The architectural design for combining LLMs with logic programs is demonstrated solely through LLM-generated logic programs, their execution via logic interpreters, and manual evaluation. This paper does not address post-training fine-tuning, adapter layers, retrieval-augmented generation, knowledge injection, or reinforcement learning, which remain areas for future work.

The experiments process only a narrow set of policies, questions, and answers in terms of data. In future work, a broader range of cases should be explored to gain deeper insights into the performance of the demonstrated approaches. Additionally, we included only a subset of available LLMs in the analysis and focused solely on Prolog as a logic interpreter. Future work should incorporate a wider variety of models and interpreters to enhance generalizability and robustness.

Regarding prompt tuning, we applied only an explicit \textit{Chain-of-logic} Prolog encoding, derived from learning on the Stanford CodeX \textit{Insurance Analyst}. Future work should explore additional encoding strategies, such as self-ask decomposition-based reasoning, iterative refinement, or reinforcement learning with thought tracing.

In this paper, we focused on accuracy and consistency measurements while qualitatively highlighting certain aspects of explainability and auditability without formal evaluation. Future work should incorporate a broader set of metrics to provide a more comprehensive assessment of the performance gains in neuro-symbolic AI designs.

\bibliography{custom}
\appendix

\onecolumn

\section{Appendix / supplemental material}

\subsection{Simplified Chubb Hospital Cash Benefit Policy}
\label{app:simplified_chubb}
\noindent Between:\\
CODEX INSURANCE LIMITED (\textquotedblleft us\textquotedblright)\\
and\\
\_\_\_\_\_\_\_\_\_\_\_\_\_\_\_\_ (\textquotedblleft You\textquotedblright)

\vspace{1em}
This policy is provided on the following terms and conditions:

POLICY IN EFFECT AND CONDITIONS

1.1 The payment of any benefit under this policy is conditioned on the policy being in effect at the time of the hospitalization for sickness or accidental injury on which the claim for such benefit is premised. The policy will be in effect if:
\begin{enumerate}
    \item This agreement is signed, 
    \item The applicable premium for the policy period has been paid, and
    \item The condition set out in Section 1.3 is still pending or has been satisfied in a timely fashion, and
    \item The policy has not been canceled.
\end{enumerate}

1.2 Cancelation will be deemed to have occurred if there is fraud, or any misrepresentation or material withholding of any information provided by you to the Company in connection with any communication or information relating to this policy, or if the condition set out in Section 1.3 has not been satisfied in a timely fashion. It will also be automatically canceled at midnight, US Eastern time then in effect, on the last day of the policy term described in Section 5 below.

1.3 No later than the 7th month anniversary of the effective date of this policy, you will supply us with written confirmation from the medical provider in question of a wellness visit for yourself with a qualified medical provider occurring no later than the 6th month anniversary of the effective date of this policy.

GENERAL EXCLUSIONS

2.1 Your policy will not apply to, and no benefit will be paid with respect to, any event causing sickness or accidental injury arising directly or indirectly out of:
\begin{enumerate}
    \item Skydiving; or
    \item Service in the military; or
    \item Service as a fire fighter; or
    \item Service in the police; or
    \item If your age at the time of the hospitalization is equal to or greater than 80 years of age.
\end{enumerate}

GENERAL CONDITIONS

3.1 Where does Your Policy apply?

3.1.1 Your Policy insures You twenty-four (24) hours a day anywhere in the world.

3.2 Arbitration

3.2.1 If any dispute or disagreement arises regarding any matter pertaining to or concerning this Policy, the dispute or disagreement must be referred to arbitration in accordance with the provisions of the Arbitration Act (Cap. 10) and any statutory modification or re-enactment thereof then in force, such arbitration to be commenced within three (3) months from the day such parties are unable to settle the dispute or difference. If You fail to commence arbitration in accordance with this clause, it is agreed that any cause of action and any right to make a claim that You have or may have against Us shall be extinguished completely. Where there is a dispute or disagreement, the issuance of a valid arbitration award shall also be a condition precedent to our liability under this Policy. In no case shall You seek to recover on this Policy before the expiration of sixty (60) days after written proof of claim has been submitted to Us in accordance with the provisions of this Policy.

3.3 Laws of New York

3.3.1 Your Policy is governed by the laws of New York.

3.4 US Currency

3.4.1 All payments by You to Us and by Us to You or someone else under your policy must be in United States currency.

3.5 Premium

3.5.1 The premium described in Section 5 below shall be paid in one lump sum at the signing of the policy.

3.6 Policy Term
The term of this policy will begin on the date accepted by Us as signified by our signature of the policy (the effective date) and will last for a period of one year from that date, unless previously canceled pursuant to Section 1 above.

\subsection{Queries and Correct Answers for Empirical Evaluation on Chubb Contract}
\label{app:queries_and_answers}

All queries are preceded by the disclaimer: \say{Assuming all other conditions are met and no other exclusions apply (where by 'other,' I mean anything not referenced in the query that follows),\ldots} \\

\textbf{Query 1: } \say{will my policy apply if I was hospitalized by burns suffered while doing my duty as a firefighter?}
\textbf{Answer: } \say{No.}
\vspace{0.1in}

\textbf{Query 2: } \say{will my policy apply if I am 78 years old at the time of hospitalization?}
\textbf{Answer: } \say{Yes.}
\vspace{0.1in}

\textbf{Query 3: } \say{will my policy apply if I was hospitalized for pneumonia 5 months after the policy's effective date, and my age at the time of hospitalization is 65?}
\textbf{Answer: } \say{Yes.}
\vspace{0.1in}

\textbf{Query 4: } \say{will my policy apply if I was hospitalized due to a fall while traveling abroad and I had given confirmation of my wellness visit 8 months after the policy's effective date?}
\textbf{Answer: } \say{No.}
\vspace{0.1in}

\textbf{Query 5: } \say{will my policy apply if I was hospitalized for punching my own face to show off for my friends and I did not commit fraud or misrepresentation?}
\textbf{Answer: } \say{No.}
\vspace{0.1in}

\textbf{Query 6: } \say{will my policy apply if I was hospitalized due to an injury sustained while skydiving, my age at the time of hospitalization was 79, and proof of my wellness visit was provided 6.5 months after the policy's effective date?}
\textbf{Answer: } \say{No.}
\vspace{0.1in}

\textbf{Query 7: } \say{will my policy apply if I was hospitalized for a heart attack, proof of the wellness visit was submitted 2 months after the policy's effective date, and my age at the time of hospitalization was 75?}
\textbf{Answer: } \say{Yes.}
\vspace{0.1in}

\textbf{Query 8: } \say{will my policy apply if I was hospitalized after being injured in a military training exercise, the hospitalization occurred within the policy term, and I did not commit fraud?}
\textbf{Answer: } \say{No.}
\vspace{0.1in}

\textbf{Query 9: } \say{will my policy apply if I was hospitalized due to my son biting me in the ankle, proof of my wellness visit was provided 6 months after the effective date, and I was serving as a police officer at the time of hospitalization?}
\textbf{Answer: } \say{Yes.}
\vspace{0.1in}

\subsection{Prompts Provided to LLMs}
\label{app:prompts_vanilla_unguided}

\subsubsection{Prompt for Vanilla LLM Approach}
\label{app:prompt_vanilla}
The following is the prompt used in the Vanilla LLM approach described in \S\ref{sec:vanilla_llm}. 

\begin{itemize}
\item [--] Below, you are provided
    \begin{enumerate}
        \item {The full text of an insurance contract} 
        \item {A specific question about whether a claim in the given scenario is covered under the terms of this insurance contract}
    \end{enumerate}
    
\item [--] Assume that the policy agreement has been signed, and the premium has been paid on time.

\item [--] Assume that all other conditions are satisfied, and no exclusions apply unless explicitly referenced in the query.

\item [--] Your task:
    \begin{enumerate}
        \item Evaluate whether the claim described in the question is covered under the insurance contract.
        \item Respond with **only** one of the following: ``Yes'', ``No'', or ``I do not know''.
        \item Do not provide any explanations or reasoning.
    \end{enumerate}

\item [--] Insurance contract: \{text\_content\}
\item [--] Question: \{query\}

\end{itemize}

\subsubsection{Prompt for Unguided Prolog Generation}
\label{app:prompt_unguided}

The following is the prompt used in \S\ref{sec:exp_unguided} to generate Chubb insurance policy encoding. 

\begin{itemize}
    \item [--] Given the insurance contract below, translate the document into valid Prolog rules so that I can run a Prolog query on the code regarding whether or not some claim is covered under the policy and receive the correct answer to the question.

    \item [--] Please fully define all predicates and DO NOT define any facts, only rules that can be used to answer queries on this insurance contract.

    \item [--] Assume that all dates/times in any query to this code (apart from the claimant's age) will be given RELATIVE to the effective date of the policy (i.e. there will never be a need to calculate the time elapsed between two dates). Take dates RELATIVE TO the effective date into account when writing this encoding.

    \item [--] Assume that the agreement has been signed and the premium has been paid (on time). There is no need to encode rules or facts for these conditions.

    \item [--] Return only Prolog code in your reply. No explanation is necessary.

    \item [--] Ensure that:
        \begin{enumerate}
            \item The legal text is appropriately translated into correct Prolog rules.
            \item The output does not redefine, misuse, or conflict with any built-in Prolog predicates.
            \item If dynamic predicates are necessary, they are declared and managed correctly.
            \item All predicates used in the generated Prolog code, including those referenced in the query, are fully defined and error-free to prevent issues like ``procedure does not exist.''
            \item Logical relationships, conditions, and dependencies in the text are faithfully represented in the Prolog rules to ensure accurate query results.
        \end{enumerate}
    \item [--] Insurance contract: \{text\_content\}
\end{itemize}

The following is the prompt used in \S\ref{sec:exp_unguided} to generate claim encodings.

\begin{itemize}
    \item [--] I have given below:
        \begin{enumerate}
            \item A question about whether or not the policy defined in a given insurance contract applies in a particular situation
            \item The text of the insurance contract
            \item A Prolog encoding of the insurance contract
        \end{enumerate}
    \item [--] Encode the question into a Prolog query such that it can be run on the given Prolog encoding of the insurance contract, returning the correct answer to the question.
    \item [--] Assume that the agreement has been signed and the premium has been paid (on time). There is no need to encode rules or facts for these conditions.
    \item [--] Return only Prolog query in your reply. No explanation is necessary.
    \item [--] Ensure that:
        \begin{enumerate}
            \item The output does not redefine, misuse, or conflict with any built-in Prolog predicates.
            \item If dynamic predicates are necessary, they are declared and managed correctly.
            \item All predicates used in the generated Prolog code, including those referenced in the query, are fully defined and error-free to prevent issues like "procedure does not exist."
            \item Logical relationships, conditions, and dependencies in the text are faithfully represented in the Prolog rules to ensure accurate query results.
            \item No absolute dates/times (apart from the claimant's age) are encoded in your query. Only include dates/times RELATIVE to the effective date of the policy (again, except for age).
            \item Set any facts/rules/parameters in the code such that ALL conditions (for the policy to apply) which are UNRELATED to the above query are satisfied.
            \item Set any facts/rules/parameters in the code such that NO exclusions (which would prevent the policy from applying) which are UNRELATED to the above query are satisfied.
        \end{enumerate}
    \item [--] Question:\{query\}
    \item [--] Insurance contract: \{text\_content\}
    \item [--] Insurance contract Prolog encoding: \{policy\_encoding\}
\end{itemize}



\subsubsection{Prompt for Generating LLM Encodings of Insurance Analyst Coverages}
\label{app:art-prompt}
\begin{itemize}
  \item I have provided below all of the text that pertains to a coverage (or section) 
        of a health insurance policy.
    \begin{itemize}
      \item The text defines all conditions and exclusions that determine 
            whether a patient's claim is covered under this coverage of the policy.
    \end{itemize}

  \item Please encode a Prolog rule, \verb|'covered(C)'|, which is true 
        exactly when the patient's claim, \verb|'C'|, is covered.
    \begin{itemize}
      \item I have also provided below a set of claim facts that will be 
            defined in any Prolog query to the \verb|'covered(C)'| rule.
        \begin{itemize}
          \item These are of the form 
                \verb|'claim_Fact(C, Value)'|, meaning they will be defined for 
                the claim \verb|'C'| by filling in a value for \verb|Value|.
          \item Under each of these facts, I have listed the valid inputs which 
                can be filled in for \verb|Value| (or in some cases, the type of 
                valid input, such as number or date) in the fact.
          \item Use ALL OF (AND ONLY) these claim facts to write your 
                Prolog encoding for the policy coverage.
        \end{itemize}

      \item Finally, I have provided below a set of supporting (pre-defined) 
            Prolog predicates.
        \begin{itemize}
          \item Under each supporting predicate, I have listed the arguments to 
                the predicate with their category (e.g.\ atom, number) and a 
                description, as well as a description of what the predicate does.
          \item Use ALL OF these supporting predicates to write your 
                Prolog encoding for the policy coverage.
        \end{itemize}
    \end{itemize}

  \item The Prolog rule that you encode, \verb|'covered(C)'|, must incorporate 
        all relevant criteria from the policy coverage.

  \item Return only Prolog code in your reply. 
        NO EXPLANATION is necessary.

\end{itemize}

\subsection{LLM Encodings of FSH Criteria in ART Coverage}
\subsubsection{GPT-4o Encoding of FSH Criteria in ART Coverage}
\begin{verbatim}
validate_day_3_fsh(C) :-
    % Get patient age at time of service
    patient_age_at_time_of_service(C, Age),
    (Age < 40 -> % For patients under 40
        most_recent_day_3_fsh_test(C, fsh_test(FSHLevel, _, _)),
        FSHLevel < 19
    ; Age >= 40 -> % For patients 40 and above
        day_3_fsh_tests_since_age_40_in_claim(C, FSHTests),
        \+ exists_test_with_level_at_least_19(FSHTests)).
\end{verbatim}
\label{app:4o-fsh}

\subsubsection{OpenAI o1 Encoding of FSH Criteria in ART Coverage}
\begin{verbatim}
meets_fsh_criteria(C) :-
    patient_age_at_time_of_service(C, Age),
    (
        ( Age < 39 ) ->
            most_recent_day_3_fsh_test(C, fsh_test(Level, _, _)),
            Level < 19
        ;
        ( Age = 39 ) ->
            claim_date_service_received(C, ServiceDate),
            most_recent_day_3_fsh_test(C, fsh_test(Level, TestDate, _)),
            leq_n_months_diff(TestDate, ServiceDate, 6),
            Level < 19
        ;
        ( Age >= 40 ) ->
            claim_date_service_received(C, ServiceDate),
            (
                claim_patient_has_premature_ovarian_failure(C, yes) ->
                    most_recent_day_3_fsh_test(C, fsh_test(Level, TestDate, _)),
                    leq_n_months_diff(TestDate, ServiceDate, 6),
                    Level < 19
                ;
                    day_3_fsh_tests_since_age_40_in_claim(C, Tests),
                    \+ exists_test_with_level_at_least_19(Tests),
                    most_recent_day_3_fsh_test(C, fsh_test(_, TestDate, _)),
                    leq_n_months_diff(TestDate, ServiceDate, 6)
            )
    ).
\end{verbatim}
\label{app:o1-fsh}

\end{document}